\documentclass[conference]{IEEEtran}%
\usepackage{amsfonts}
\usepackage{amssymb}
\usepackage{amsmath}
\usepackage{graphicx}
\usepackage{color}%
\providecommand{\U}[1]{\protect\rule{.1in}{.1in}}
\newtheorem{theorem}{Theorem}

\newtheorem{lemma}{Lemma}

\begin{document}


\title{Delayed Channel State Information: Incremental Redundancy with Backtrack Retransmission}
\author{\authorblockN{Petar Popovski}\\
\authorblockA{Department of Electronic Systems, Aalborg University\\
Email: petarp@es.aau.dk}}
\maketitle

\begin{abstract}
In many practical wireless systems, the Signal-to-Interference-and-Noise Ratio (SINR) that is applicable to a certain transmission, referred to as Channel State Information (CSI), can only be learned \emph{after} the transmission has taken place and is thereby outdated (delayed). For example, this occurs under intermittent interference. 
We devise the \emph{backward retransmission (BRQ)} scheme, which uses the delayed CSIT to send the optimal amount of incremental redundancy (IR). BRQ uses fixed-length packets, fixed-rate $R$ transmission codebook, and operates as Markov block coding, where the correlation between the adjacent packets depends on the amount of IR parity bits. When the delayed CSIT is full and $R$ grows asymptotically, the average throughput of BRQ becomes equal to the value achieved with prior CSIT and a fixed-power transmitter; however, at the expense of increased delay. The second contribution is a method for employing BRQ when a limited number of feedback bits is available to report the delayed CSIT. The main novelty is the idea to assemble multiple feedback opportunities and report multiple SINRs through vector quantization. This challenges the conventional wisdom in ARQ protocols where feedback bits are used to only quantize the CSIT of the immediate previous transmission. 
\end{abstract}

\section{Introduction}
\subsection{Motivation}

Channel State Information at the Transmitter (CSIT) is essential for achieving high spectral efficiency in wireless  systems. Ideally, CSIT should be known before the transmission has started, as in that case the transmitter can optimize the parameters of the transmission, such as the power used or the precoding that is applied in case of Multiple Input Multiple Output (MIMO) transmission. In Frequency Division Duplex (FDD) systems, provision of CSIT necessarily happens through a feedback from the receiver. On the other hand, Time Division Duplex (TDD) systems can utilize channel reciprocity and the transmitter can obtain CSIT by measuring a signal received in the opposite direction. 

\begin{figure}[t]
  \centering
   \includegraphics{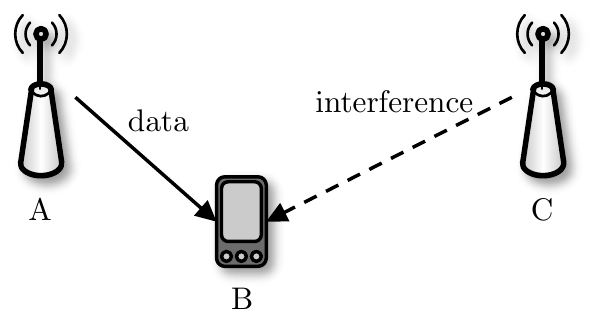}
\caption{An example of communication scenario in which it is not possible to know the conditions at the receiver $B$ before the transmission from $A$ takes place use to intermittent interference from $C$.}\label{fig:AtoBinterfC}
\end{figure}

Reciprocity loses its utility when there is interference. Consider the scenario on Fig.~\ref{fig:AtoBinterfC}, where $A$ is the transmitter, $B$ is the receiver and $C$ is an interferer that affects $B$. By assuming that $A$ knows the noise power at $B$, channel reciprocity enables $A$ to estimate the Signal-to-Noise Ratio (SNR), denoted $\gamma_{AB}$, at which $B$ receives the signal of $A$. When the interferer $C$ is active, the quantity that is relevant for the transmitter is the Signal-to-Interference-and-Noise Ratio (SINR) and $A$ has no way of knowing it unless $B$ explicitly reports either the Interference-to-Noise Ratio (INR) $\gamma_{CB}$ or the actual SINR. Hence, the notion of CSIT is broader and incorporates knowledge about the exact conditions at which the signal has been received, including the interference. In many practical scenarios, the interference is intermittent and $B$ cannot know whether/which interferer will be active when the actual transmission of $A$ takes place. Furthermore, even if there is no interference and the channel is TDD, there could not have been a reliable way to measure the SNR $\gamma_{AB}$ before the transmission, for example due to the scheduling structure in the system or fast channel dynamics. 
On the other hand, \emph{after} $A$ has completed its transmission, $B$ can send feedback to $A$ about what the actual received SNR or SINR had been during that transmission. We refer to this as a \emph{posterior CSIT} or \emph{delayed CSIT}, as opposed to \emph{prior CSIT}, known before the actual transmission. 

The performance of the communication link that operates with posterior CSIT is certainly inferior compared to the link that operates with prior CSIT, as there are certain parameters that cannot be optimally adapted if CSIT is not known before the transmission. 
If CSIT is not available before the transmission, then the power that $A$ uses for transmission cannot follow the water filling principle and thus leads to \emph{irrecoverable loss}, since at the time of getting CSIT, the power has already been used. In other words, if the channel gain is below the threshold, then the power is irreversibly lost to a bad channel state. Similarly, the lack of prior CSIT in MIMO transmission prevents the sender to have optimal spatial focusing and power is irreversibly lost in spatial dimensions that are not optimal. As the last example, \emph{multiuser diversity} is not recoverable when the posterior CSIT is available, because prior CSIT is used to select the user with the best channel conditions and transmit data to her. When the user is selected blindly, the power may be irreversibly lost to a ``bad'' user.

In this paper we introduce \emph{backtrack retransmission (BRQ)}, capable to use the delayed CSIT in a way that approaches the average throughput achieved with prior CSIT and fixed transmission power. A prominent feature of BRQ is that it adapts the data rate using a fixed rate$-R$ codebook at the physical layer and fixed packet length, but only changing the amount of parity bits based on the delayed CSIT. Thus, BRQ operates as Markov block coding, where the correlation between the adjacent packets depends on the amount of IR parity bits. By selecting asymptotically high rate $R$ and when the delayed CSIT is full, the average throughput becomes equal to the one with prior CSIT. However, this happens at the expense of asymptotically increased delay. The extension of BRQ to the case with a finite number of feedback bits gives rise to a novel transmission scheme: instead of sending feedback after each packet, the feedback bits from multiple slots are assembled and, using vector quantization, provide information relevant to a number of packets transmitted in the past. This challenges the conventional wisdom in ARQ protocols, in which the content of the feedback bits is only associated with the transmission immediately preceding those bits. The price paid by the BRQ is in the increased delay. 

\subsection{Related Work}

The problem of HARQ over block fading channels has been addressed in many works, but two recent works stand out in their relation to the current work, \cite{Tuninetti11} and \cite{Szczecinski13}, respectively. The scenario considered in \cite{Tuninetti11} is essentially different from our scenario of interest, as the CSIT in is not delayed, but it arrives through a finite number of bits as a prior CSIT, such that the transmitter can adapt the power. The reference \cite{Tuninetti11} provides good insights on the role of the feedback as a CSIT quantizer, but the feedback bits are associated only with the immediately preceding transmission and it is not allowed to accumulate feedback bits over consecutive slots, as we do in the quantized scheme in Section~\ref{sec:Quant_BRQ}. The scenario considered in \cite{Szczecinski13} is directly related to the one treated in this paper, as the authors assume that at the time of transmission, the received CSIT is not correlated with the actual conditions on the channel. The IR bits in \cite{Szczecinski13} are sent in a standalone way, using variable-length packets --- the authors of \cite{Szczecinski13} correctly point out that this could be a problem in multiuser systems and therefore they propose a workaround by grouping multiple transmissions into fixed-length resources. On the other hand, our approach works with fixed-length packets/slots at the physical layer, such that it naturally follows the modus operandi of the contemporary wireless systems, such as e. g. LTE~\cite{LTE}. The method in \cite{Szczecinski13} uses rate adaptation at the physical layer, while in our case the physical layer parameters are fixed and a single combination of coding/modulation of rate $R$ is applied. Regarding the implementation of IR, both \cite{Tuninetti11} and \cite{Szczecinski13} describe it implicitly, through sufficient accumulated mutual information, while our scheme is based on explicit method for creating the IR bits through random binning. The approach in \cite{Szczecinski13} gives rise to a complex optimization problem for determining the IR bits and the rate, while in our method this is straightforwardly based on the mechanism of random binning. Finally, we extend our method with delayed CSIT to the case with finite-bit feedback, while  \cite{Szczecinski13} treats only the case of full delayed CSIT. 

We remark here that the issue of delayed CSIT has recently sparkled a significant interest in terms of improvements that can be achieved in terms of Degrees of Freedom at high SNRs ~\cite{Tse12}~\cite{Tandon12}. However, that line of work does not deal with HARQ protocols and is directed towards multi-user scenarios, while in our case we consider relatively low SNR and a single link. 

\section{System Model and Illustrative Examples}

\subsection{System Model}

We consider a single-user channel with block fading and Gaussian noise. From this point on we use the term SNR, as we will not explicitly consider interference and attribute the channel variation to fading, but we keep in mind that the concepts are applicable to SINR. TX transmits to RX in slots, each slot contains a full packet and takes $N$ channel uses. The value of $N$ is fixed, unless stated otherwise, and sufficiently large to offer reliable communication that is optimal in an information-theoretic sense. The complex $N-$dimensional received signal vector in slot $t$ is:
\begin{equation}\label{eq:ReceivedSignal}
\mathbf{y}_t=\sqrt{\gamma_t} \mathbf{x}_t+\mathbf{z}_t
\end{equation}
where $\mathbf{x}_t$ is the transmitted signal normalized to have unit power and $\mathbf{z}_t$ is a complex random vector with unit variance that represents the contribution of the noise and the interference in slot $t$. 
Hence, the SNR in the $t-$th slot is equal to $\gamma_t$, drawn independently from a probability density $p_{\Gamma}(\cdot)$ and is unknown to TX prior to the transmission. We note that (\ref{eq:ReceivedSignal}) assumes a fixed transmission power, which is reasonable considering the fact that we always assume that  $\gamma_t$ is unknown when the transmission takes place; more elaboration will follow. 

TX uses a single code/modulation combination of rate $R$ and thus sends a total of $b=NR$ data bits in the packet. The maximal rate that the channel TX-RX can support in slot $t$ is:
\begin{equation} \label{eq:Cgamma}
C(\gamma_t)=\log_2(1+\gamma_t) \qquad \textrm{[bits/c. u.]}
\end{equation}
If $R \leq C(\gamma)$ then RX receives the packet correctly, otherwise an outage occurs. Upon outage, RX buffers the received signal in order to use it in future decoding attempts. An efficient way to treat the outage is to use \emph{incremental redundancy (IR)} \cite{Tuninetti11}: TX sends additional $r<b$ parity bits to RX, which RX can combine with the signal received during the previous $N$ channel uses and thus possibly recover the original packet. If TX uses $M$ channel uses to transmits the $r$ additional bits and eventually the packet is decoded correctly, then the data rate achieved between TX and RX is $R_r=\frac{N}{N+M}R$. Note that in general $M \neq N$. An \emph{optimal} incremental redundancy would select the retransmitted bits in such a way that if the accumulated mutual information at the receiver becomes sufficient, the signal is decoded \cite{Tuninetti11} \cite{Szczecinski13}. The interesting question is how is $R_r$ related to the channel capacity $C(\gamma)$?

\subsection{Two Examples of Incremental Redundancy}

For the first example, TX sends the packet using a slot with $N$ channel uses and learns the SNR $\gamma$ after the slot. The value of $\gamma$ is such that 
$R>C(\gamma)$ and an outage occurs. For this example we deviate from our model in two ways: (1) we temporarily break away from the fixed slot structure and assume that the retransmission from TX can take $M$ channel uses, where $M \neq N$ (as in \cite{Szczecinski13}); (2) the SNR during those $M$ channel uses is constant and equal to $\gamma$, such that TX can perfectly choose the transmission rate used to send the IR bits without experiencing outage. By knowing $\gamma$, TX knows that RX has a side information about the transmitted packet that amounts to $NC(\gamma)<NR=b$ bits, such that the minimal amount of parity bits that $A$ should provide to $B$ in order to decode the packet is:
\begin{equation}\label{eq:r_bits}
r=N(R-C(\gamma))
\end{equation}
The \emph{operational interpretation} of the minimal number of parity bits bears resemblance to \emph{random binning}, introduced in the Slepian-Wolf problem of distributed source coding~\cite{Kramer07}. In the Slepian-Wolf problem, the amount of information that one node sends depends on the knowledge of the side information that the other node provides to the receiver. In our setting with a retransmission, TX can act as two different senders separated in time: after the first transmission, TX learns from the CSIT the amount of side information available at RX, adjusts the amount of parity bits and sends them to RX. Specifically,  TX divides the $2^b$ messages into $2^r$ bins, where the number of bins is adjusted to the received CSIT, and X sends to RX $r$ bits to describe to which bin the previous message belongs to. If RX decodes the $r$ bits correctly, it combines the bin information with the side information present at RX from the first $N$ channel uses and, RX can decode the original $b$ bits correctly almost surely as $N \rightarrow \infty$.

The rate at which TX sends the redundancy bits is $R^{\prime}=C(\gamma)$, such that number of channel uses for retransmission is:
\begin{equation}
M=\frac{r}{R^{\prime}}=\frac{N(R-C(\gamma))}{C(\gamma)}
\end{equation}
and the equivalent rate achieved from TX to RX is:
\begin{equation}
R_{TX-RX}=\frac{b}{N+M}=\frac{NR}{N+\frac{N(R-C(\gamma))}{C(\gamma)}}=C(\gamma)
\end{equation}
which is the same as if TX knew the CSIT a priori. We can conclude that, if the channel is constant during the original rate-$R$ transmission and the incremental redundancy transmission, but TX can only learn the CSIT after the first transmission, the achievable rate between TX and RX is:
\begin{equation}
R_{AB}=\min \{R, C(\gamma) \}
\end{equation}
Besides the inability to adapt the power, the penalty for not knowing the prior CSIT can occur due to a low value of $R$, i.~e. TX cannot take advantage of the very high SNRs. 

\begin{figure}[t]
  \centering
   \includegraphics{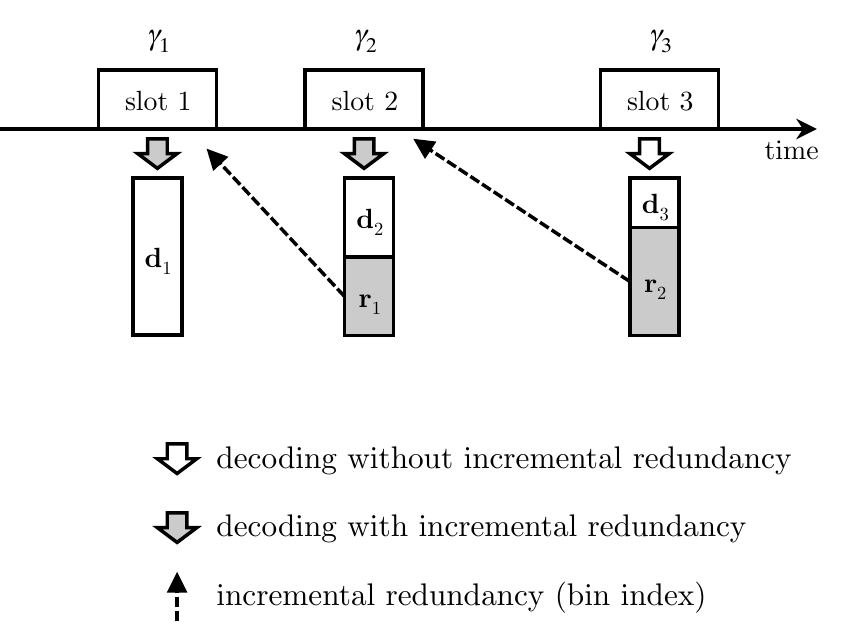}
\caption{An example of backtrack decoding with posterior CSIT. In the $i-$th slot, $\mathbf{d}_i$ are the new bits and $\mathbf{r}_{i-1}$ are the redundancy bits, used as a bin index to decode the transmission in slot $i-1$.}\label{fig:Backtracking3slots}
\end{figure}

But a careful reader can object to the consistency of the previous example. First, to send the IR bits, TX uses a prior CSIT, such that the example cannot cover e. g. a scenario with intermittent interferer. Knowing CSIT in advance also opens the possibility for employing power control, as in \cite{Tuninetti11}. On the other hand, we persist to the assumption that the channel must be unknown at the time of transmission and ask whether we can still recover the rate. Furthermore, the example violates the required fixed-slot structure and a single coding/modulation combination with a fixed rate.

Therefore, as a second example assume that TX sends to TX in slots with fixed size of $N$ channel uses. The SNR from slot to slot is uncorrelated and available to TX only after the transmission. Let TX always apply the same transmission rate $R$ and let us observe three consecutive slots with SNRs equal to $\gamma_1, \gamma_2, \gamma_3$. This is illustrated on Fig.~\ref{fig:Backtracking3slots}. We deliberately select the SNRs to satisfy
\begin{equation}
C(\gamma_1), C(\gamma_2) <R \qquad C(\gamma_3)>R 
\end{equation}
such that outage occurs in the first two slots and, eventually, the transmission of TX can be decoded in the third slot. TX transmits $b=NR$ new bits in the first slot, denoted by $\mathbf{d}_1$. For the second slot, TX knows $\gamma_1$ and sends again $b=NR$ data bits, but the message is prepared in the following way:
\begin{itemize}
\item The first $r_1=N(R-C(\gamma_1))$ bits are parity bits, denoted by $\mathbf{r}_1$, used to describe the bin index that can be combined with the signal from slot 1 to recover the bits $\mathbf{d}_1$.
\item The remaining $b-r_1=NC(\gamma_1)$ bits, denoted by $\mathbf{d}_2$ are \emph{new bits} transmitted in the second slot. 
\end{itemize}
It must be noted that the parity bits and the new bits are only separable in a digital domain, but not at the physical layer i.~e. the whole packet, sent at rate $R$, needs to be decoded correctly and then RX extracts the parity bits and the new bits. The insertion of parity bits creates dependency between two adjacent packet transmissions, such that the scheme effectively applies Markov block coding. 

Coming back to the example, outage also occurs in the second slot, such that the message in the third slot consists of $r_2=N(R-C(\gamma_2))$ redundancy bits $\mathbf{r}_2$ and $d_3=NC(\gamma_2)$ new bits $\mathbf{d}_3$. Note that, according to the illustration on Fig.~\ref{fig:Backtracking3slots}, in the example $r_1<r_2$, which means that $\gamma_2<\gamma_1$. As $C(\gamma_3)>R$, RX decodes the message slot 3 and recovers $\mathbf{d}_3$ and $\mathbf{r}_2$. It then uses $\mathbf{r}_2$ as incremental redundancy to decode the transmission in slot 2 and recover $\mathbf{d_2}$ and $\mathbf{r}_1$. Finally, RX uses $\mathbf{r}_1$ to decode the transmission from slot 1 and recover $\mathbf{d}_1$. 

The average rate is $\bar{R}=\frac{b_1+b_2+b_3}{3N}$, which results in
\begin{equation}
\bar{R}=\frac{R+C(\gamma_1)+C(\gamma_2)}{3}
\end{equation}
The average rate that could have been achieved over the three slots if the CSIT were known a priori is: 
\begin{equation}
\bar{R}_{\mathrm{prior}}=\frac{C(\gamma_3)+C(\gamma_1)+C(\gamma_2)}{3}
\end{equation}
Again, a loss compared to prior CSIT can occur due to low $R$, but otherwise the data rates that are achievable with prior CSIT can be recovered with delayed CSIT. 

The second example illustrates the central proposal in this paper, IR with \emph{backtrack retransmission (BRQ)}. We note that power adaptation cannot help under the assumption that the SNR $\gamma_t$ is unknown and uncorrelated with $\gamma_{t-1}$ at the time of transmission and here is a sketch of the argument to support this. Referring to Fig.~\ref{fig:Backtracking3slots}, it should be noted that RX does not have any side information about the bits $\mathbf{r}_1$ and $\mathbf{r}_2$, since they represent the minimal amount of information that needs to be retransmitted. Therefore, the information carried in $\mathbf{r}_1$ and $\mathbf{r}_2$ is new in the same sense that the information in $\mathbf{d}_1,\mathbf{d}_2, \mathbf{d}_3$ is new. Hence, each transmitted packet carries equal amount of $NR$ new information bits and, since the channel is not known in advance and for symmetry reasons, the transmission power should be equal in each slot.

\section{Backtrack Retransmission (BRQ) with Full CSIT}


Here we specify the BRQ protocol with full CSIT. TX sends to RX using a codebook of rate $R$ and the minimal SNR required to decode it is:
\begin{equation}
\gamma_R=C^{-1}(R)=2^R-1
\end{equation}
We pick the system at slot $t$ that has SNR of $\gamma_t$, unknown to TX prior to the transmission. At the end of the slot $t-1$, RX sends the value $\gamma_{t-1}$ to the TX, such that when slot $t$ starts, both TX and RX know $\gamma_{t-1}, \gamma_{t-2}, \ldots$ In addition, RX knows $\gamma_t$, as we assume coherent reception. It is assumed that the last slot in which the receiver has successfully decoded the packet is $t-L$, where $L>1$ such that $\gamma_{t-L} \geq \gamma_R$ and $\gamma_{t-l} <\gamma_R$ for $l=2, 3, \ldots L-1$. Operation in slot $t$:
\renewcommand*\theenumi{TX-\arabic{enumi}}
\renewcommand*\labelenumi{\theenumi)}
\begin{itemize}
\item Transmitter side:
\begin{enumerate}
\item  Receive the value of $\gamma_{t-1}$.
\item If $\gamma_{t-1} \geq \gamma_R$, fetch $NR$ new bits and transmit them at rate $R$. 
\item  Else $\gamma_{t-1} < \gamma_R$ and create the bin, consisting of $N(R-C(\gamma_{t-1}))$ parity bits and fetch $NC(\gamma_{t-1})$ new bits. Concatenate the parity and the new bits into a $NR-$bit packet and send at rate $R$.
\end{enumerate}
\item Receiver side:
\renewcommand*\theenumi{RX-\arabic{enumi}}
\renewcommand*\labelenumi{\theenumi)}
\begin{enumerate}
\item If $\gamma_{t-1} < \gamma_R$, store the received signal $\mathbf{y}_t$ in memory for later use.
\item Else, decode the packet, set $l=0$ and while $l<L$ do the following:
\begin{enumerate}
\item From the packet decoded in slot $(t-l)$ extract the parity bits and the new bits. 
\item Use the parity bits from slot $(t-l)$ and the stored $\mathbf{y}_{t-l-1}$ to successfully decode the transmission of TX from slot $(t-l-1)$;   
\item Set $l=l+1$.
\end{enumerate}
\item  Send $\gamma_{t}$ to TX.
\end{enumerate}
\end{itemize}
Note that whenever TX receives feedback $\gamma_{t-1}>\gamma_R$, it is treated as an ACK. In the BRQ protocol, the decoder buffers the received signals until a slot with $\gamma_{t-1}>\gamma_R$ and then decodes all stored transmissions. The stochastic behavior of the protocol can be described by a renewal reward process~\cite{Tuninetti11}, in which a renewal occurs in a slot with $\gamma_{t-1}>\gamma_R$. The reward is calculated as the total number of information bits, excluding the parity bits, that RX decodes in the slot at which a renewal occurs. 
\begin{lemma}
Let $\gamma_{t-L}, \gamma_t \geq \gamma_R$ with $L>0$ and $\gamma_{t-l}<\gamma_R$ for $0<l<L$. Then the reward in slot $t$ is:
\begin{equation} \label{eq:RewardBits}
\rho_t=N \left(R+\sum_{l=1}^{L-1} C(\gamma_{t-l}) \right)
\end{equation}
where $N$ is the number of channel uses per slot. 
\end{lemma}
\begin{IEEEproof}
Since $\gamma_{t-l}<\gamma_R$ for $0<l<L$, from step TX-3 of the protocol it follows that the packets transmitted in slot $(t-j)$ for $0 \leq j <L-1$ will contain parity bits and new bits. The number of parity bits sent in slot $(t-j)$ is $N[R-C(\gamma_{t-j-1})]$, such that the number of new bits in slot $(t-j)$ is $NC(\gamma_{t-j-1})$. Summing up the new bits in the slots $t-L+1, t-L+2, \ldots, t$ results in~(\ref{eq:RewardBits}).
\end{IEEEproof}

\begin{theorem}
Let TX transmit data to RX using a fixed rate $R$, with full CSIT available a posteriori and using BRQ. The SNR in each slot is drawn independently from $p_{\Gamma}(\gamma)$. Then the average rate is:
\begin{equation} \label{eq:AverageRateFullCSIT}
\bar{R}=\int_{0}^{\gamma_R} p_{\Gamma}(\gamma) C(\gamma) \mathrm{d}\gamma + R \int_{\gamma_R}^{\infty} p_{\Gamma}(\gamma) \mathrm{d}\gamma 
\end{equation}
\end{theorem}
\begin{IEEEproof}
Let us consider another system, termed $R-$limited protocol, in which TX knows $\gamma_t$ at the start of the slot $t$ and uses a fixed transmission power. The $R-$limited protocol adapts the rate in the following way:  
\begin{equation}
R_t= \max \{ C(\gamma_t), R \}
\end{equation}
i.~e. it uses the instantaneous channel rate if $\gamma_t<\gamma_R$ and otherwise rate $R$. The average rate of the $R-$limited protocol is straightforwardly given by~(\ref{eq:AverageRateFullCSIT}). 

We observe the performance of the $R-$limited protocol and BRQ with full CSI on the same set of $L+1$ consecutive slots. The SNRs are selected as $\gamma_{t-L}, \gamma_t \geq \gamma_R$ and $\gamma_{t-l}<\gamma_R$ for all $l=1 \ldots L-1$ i. e. these are slots between two successful decode events for the BRQ protocol. From Lemma 1 it follows that the reward collected for slots $t-l$, with $l=0 \ldots L-1$ is identical for BRQ and for the $R-$limited protocol. Since this is valid for any set of SNRs between two decode events in BRQ, the average rate of the BRQ protocol and the $R-$limited protocol are identical, which proves the theorem. 
\end{IEEEproof}

An interesting feature of BRQ is that it adapts the rate by using a single transmission codebook, while the $R-$limited protocol needs to apply a different codebook for different $\gamma$. The adaptation in BRQ is done in digital domain, through the adaptive number of parity bits. Therefore, between two renewals, BRQ operates as Markov block coding in which the statistical dependence between the packets in slots $l$ and $l+1$ depends on the posterior CSIT received for slot $l$.

However, BRQ is able to recover the rate achievable by the prior CSIT at the expense of increased delay. We define the delay of a bit that is transmitted as a new bit in slot $t$ and decoded in slot $t+L$ to be $L$ slots. We do not define delay for a parity bit. For example, in slot $t$ TX sends $NC(\gamma_{t-1})$ new bits and $N[R-C(\gamma_{t-1})]$ parity bits. If $\gamma_{t+l} < \gamma_R$ for $0 \leq l <L$ and $\gamma_L \geq \gamma_R$, then the delay for the new bits sent in slot $t$ is $L$. On the other hand, the delay for all the bits sent in the $R-$limited adaptation protocol is zero. In order to calculate the average delay, let us define:
\begin{equation}
p_R=\int_{\gamma_R}^{\infty} p_{\Gamma}(\gamma) \mathrm{d}\gamma
\end{equation}
The time that a given bit spends in the system has a geometric distribution and the average delay is given by:
\begin{equation}
\bar{\tau}=\frac{p_R}{1-p_R} \qquad \textrm{[slots]}
\end{equation}
and increases with $R$. 

\section{BRQ with Quantized CSIT}
\label{sec:Quant_BRQ}


In this section we investigate how to use the idea of backtrack retransmission when a finite number of $F$ bits are available after each slot. We devise a strategy that sacrifices the delay performance in order to efficiently use the feedback bits and enable backtrack retransmissions. Note that if the $F$ bits that follow the $t-$th slot are used to report a quantized value of $\gamma_t$, then this is a scalar quantization. Our approach is to assemble $LF$ bits and jointly report the quantized versions of $\gamma_t, \gamma_{t+1}, \ldots \gamma_{t+L-1}$ after the slot $(t+L-1)$.

\begin{figure}[t]
  \centering
   \includegraphics{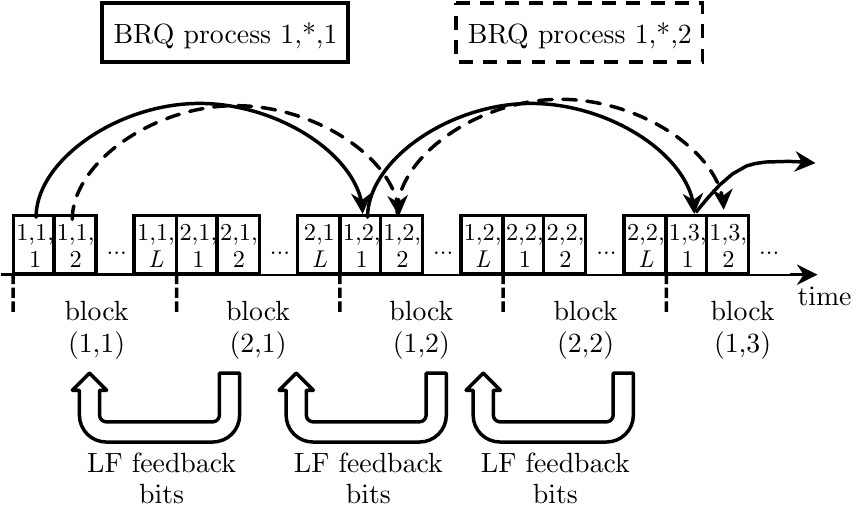}
\caption{Backtrack retransmission with vector quantization of CSIT. The LF feedback bits in the even (odd) block are used to report the SNRs in the previous odd (even) block. There are $2L$ BRQ processes running in parallel, each associated with one of the $L$ even or odd slots. Two processes are illustrated, associated with the odd slots $1$ and $2$, respectively.}\label{fig:BlockVQ}
\end{figure}

The transmission strategy can be specified as follows. We group the transmission slots into blocks of $L$ slots, where $L$ is sufficiently large. We then differentiate between odd and even blocks, such that $(1,i)$ is the $i-$th odd block and $(2,j)$ is the $j-$th even block. The $l-$th slot of an odd (even) block will be referred to as $l-$th odd (even) slot. The odd and the even blocks are interleaved in time, such that the sequence is $(1,1), (2,1), (1,2), (2,2), (3,1), \ldots$, see Fig.~\ref{fig:BlockVQ}. Let us assume that the communication starts in block $(1,1)$ and TX transmits new data bits in each slot of the block $(1,1)$, such that in total $LNR$ new bits are transmitted. The $LF$ feedback bits of block $(1,1)$ are not used, which is a waste that becomes negligible asymptotically, as the number of observed blocks goes to infinity. RX records the SNRs of each of the $L$ slots during block $(1,1)$. Specifically, we denote by $\gamma_{1,i,l}$ the SNR in the $l-$th slot of the $i-$th odd block. At the end of block $(1,1)$, $B$ performs vector quantization of $\gamma_{1,1,1}, \gamma_{1,1,2}, \cdots \gamma_{1,1,L}$ and uses the $LF$ feedback bits to report two different messages: (a) in which slots $l$ of block $(1,1)$ the decoding was successful, i.~e. $\gamma_{1,1,l} \geq \gamma_R$, and (b) the quantized SNRs $\hat{\gamma}_{1,1,1}, \hat{\gamma}_{1,1,2}, \cdots \hat{\gamma}_{1,1,L}$. These bits are transmitted to TX during the $L$ feedback opportunities of the even block $(2,1)$, as denoted on Fig.~\ref{fig:BlockVQ}. TX recovers the distorted versions of the SNRs. If TX learns that $\gamma_{1,1,l} \geq \gamma_R$, TX sends $NR$ new bits in slot $(1,2,l)$. Otherwise, $\gamma_{1,1,l} < \gamma_R$ and TX prepares $NR_{1,1,l}$ parity bits and $NR-NR_{1,1,l}$ new bits and transmits them during the slot $(1,2,l)$. The choice of $R_{1,1,l}$ depends on the received value $\hat{\gamma}_{1,1,l}$ as well as the fidelity criterion used for quantization, as explained below. During the first even block $(2,1)$, TX sends new bits in all $L$ slots, unrelated to the transmissions in block $(1,1)$. 

From the description above it can be inferred that TX runs $2L$ instances of BRQ in parallel, for the $L$ even and the $L$ odd slots, respectively. For example, the BRQ process denoted by $(1,*,l)$ is associated with the $l-$th odd slot. Let us observe the $l-$the odd slot $(1,*,l)$ and assume successful decoding occurs in block $j$ (slot  $(1,j,l)$) and the next one in block $j+J$ (slot $(1,j+J,l)$). Then, using backtrack, RX decodes the signals received in slots $(1,j+1,l), (1,j+2,l), \cdots (1,J,l)$. The operation is analogous for the even slots and the BRQ operation within the $l$ even/odd slots proceeds according to the description in the previous section. 

The key to this operation is how to perform the quantization. By definition, each SNR $\gamma_l$ is non-negative, and for the quantization we put the following fidelity criterion:
\begin{equation}\label{eq:QuantCriterion}
\gamma_l \geq \hat{\gamma}_l-d
\end{equation}
where $d$ is a positive constant distortion value. This is a rather heuristic criterion, while we will briefly address the problem of optimal criterion in Section~\ref{sec:Discussion}. The motivation for (\ref{eq:QuantCriterion}) can be explained as follows. When TX observes $\hat{\gamma}_l$ it knows that this is not the correct value, but it is desirable to know a lower bound on the true $\gamma_l$, such that TX can be sure that the amount of parity bits sent will be equal or larger than what is minimally required to recover the failed transmission in slot $l$. Thus, the number of parity bits that TX sends to RX for recovering the transmission in the $l-$th slot is: 
\begin{equation}
N(R-C(\hat{\gamma}_l-d)^+)
\end{equation}
where $(x)^+=\max\{x,0\}$. With a slight abuse of the notation, we denote the next slot of the BRQ process by $(l+1)$, such that the amount of new bits sent in the $(l+1)-$th slot is: 
\begin{equation}
NC((\hat{\gamma}_l-d)^+)
\end{equation}
If we look at a single BRQ instance, then we can use the analysis of the previous section, such that the reward in the slot $t$ in which decoding occurs (Lemma 1) can be written as:
\begin{equation}\label{eq:rho_t}
\rho_t = N \left(R+\sum_{l=1}^{L-1} C((\hat{\gamma}_{t-l}-d)^+) \right)
\end{equation}
Let $p(\hat{\gamma}|\gamma)$ be a conditional probability distribution used for quantization that satisfies the fidelity criterion~(\ref{eq:QuantCriterion}). Recall that we have designed the feedback to tell to TX in which slots decoding has occurred i.~e.  $\gamma \geq \gamma_R$, while for the remaining slots TX decides the number of parity bits based on $\hat{\gamma}$. 

The probability that a transmission is decoded is $p_R$ and the decoding events from slot to slot are independent. For sufficiently large block length $L$, the number of bits required to describe the slots in the block in which decoding occurred is $LH(p_R)$, where $H(\cdot)$ is the entropy function. Hence, the number of bits available for the vector-quantized versions of the SNRs occurring in a block is $L(F-H(p_R))$ or $F-H(p_R)$ bits per slot i.~e. SNR value. 

The probability density function $q_{R}(\gamma)$ that should be used for quantization is different from the original distribution of the SNR $p_{\Gamma}(\gamma)$, as we only need to quantize the values $\gamma < \gamma_R$. Note that $q_{R}(\gamma)$  depends on the choice of $R$.
Specifically:
\begin{equation}
q_{\Gamma}(\gamma)=\frac{p_{\Gamma}(\gamma)}{1-p_R} \mathbf{I}(\gamma < \gamma_R)
\end{equation}
where the indicator function $\mathbf{I}(x<y)=1$ if $x<y$ and is $0$ otherwise. We denote by $S_R(d)$ the rate distortion function computed for $q_{R}(\gamma)$ and using the fidelity criterion~(\ref{eq:QuantCriterion}), we can establish the following relation:
\begin{equation}\label{eq:exp_dist_bound}
d=S_R^{-1}(F-H(p_R))
\end{equation}

Using similar reasoning as in Theorem 1, considering that the $2L$ BRQ instances are statistically equal and over a long period the waste of the unused feedback in the first block disappears, we can state the following:
\begin{theorem}
Let TX transmit data to RX using a fixed rate $R$. Let there be $F$ feedback bits per slot and RX assembles the feedback bits of $L$ blocks. Part of the $LF$ bits are used to report in which slots there was a decoding event and the remaining bits are used to report quantized values of the SNRs to TX. Let $p(\hat{\gamma}|\gamma)$ be a conditional distribution that satisfies the fidelity criterion~(\ref{eq:QuantCriterion}) and $p_{\Gamma}(\gamma)$ be the density of SNR in each slot. Then the average rate is:
\begin{equation} \label{eq:AverageRateQuantCSIT}
\bar{R}=\int_{\gamma=0}^{\gamma_R} \int_{\hat{\gamma}} p(\hat{\gamma}|\gamma) p_{\Gamma}(\gamma) C((\hat{\gamma}_l-d)^+) \mathrm{d} \hat{\gamma} \mathrm{d} \gamma + R \int_{\gamma_R}^{\infty} p_{\Gamma}(\gamma) \mathrm{d}\gamma 
\end{equation}
where $d=S_R^{-1}(F-H(p_R))$.
\end{theorem}

As an example, we can consider the Rayleigh fading for which $p_{\Gamma}(\gamma)=\frac{1}{\Gamma}e^{-\frac{\gamma}{\Gamma}}$. Quantization of the exponential distribution with fidelity criterion~(\ref{eq:QuantCriterion}) has been considered in~\cite{Verdu96}.
The derivation of the rate distortion function $S_R$ for given $R$ is a problem on its own and not of direct interest in this initial paper. Instead, we use the $F-H(p_R)$ bits per SNR in a suboptimal way, making a vector quantization according to $p_{\Gamma}(\gamma)$. This is clearly suboptimal as RX reports to TX also quantized versions of the SNR values $\gamma > \gamma_R$, which is redundant. Therefore, the distortion computed accordion to the result from~\cite{Verdu96} represent an upper bound on the distortion that can be achieved if we (properly) quantize according to $q_R(\cdot)$: 
\begin{equation}\label{eq:UpperBoundDist}
d \leq \Gamma \cdot 2^{-(F-H(p_R))}
\end{equation}
As $R$ increases, $p_R$ and $H(p_R)$ decrease, while $q_R(\cdot)$ becomes  better approximated by $p_{\Gamma}(\cdot)$, such that the upper bound in~(\ref{eq:UpperBoundDist}) becomes tight. 

\section{Numerical Illustration}

\begin{figure}[t]
  \centering
   \includegraphics[width=8.3cm]{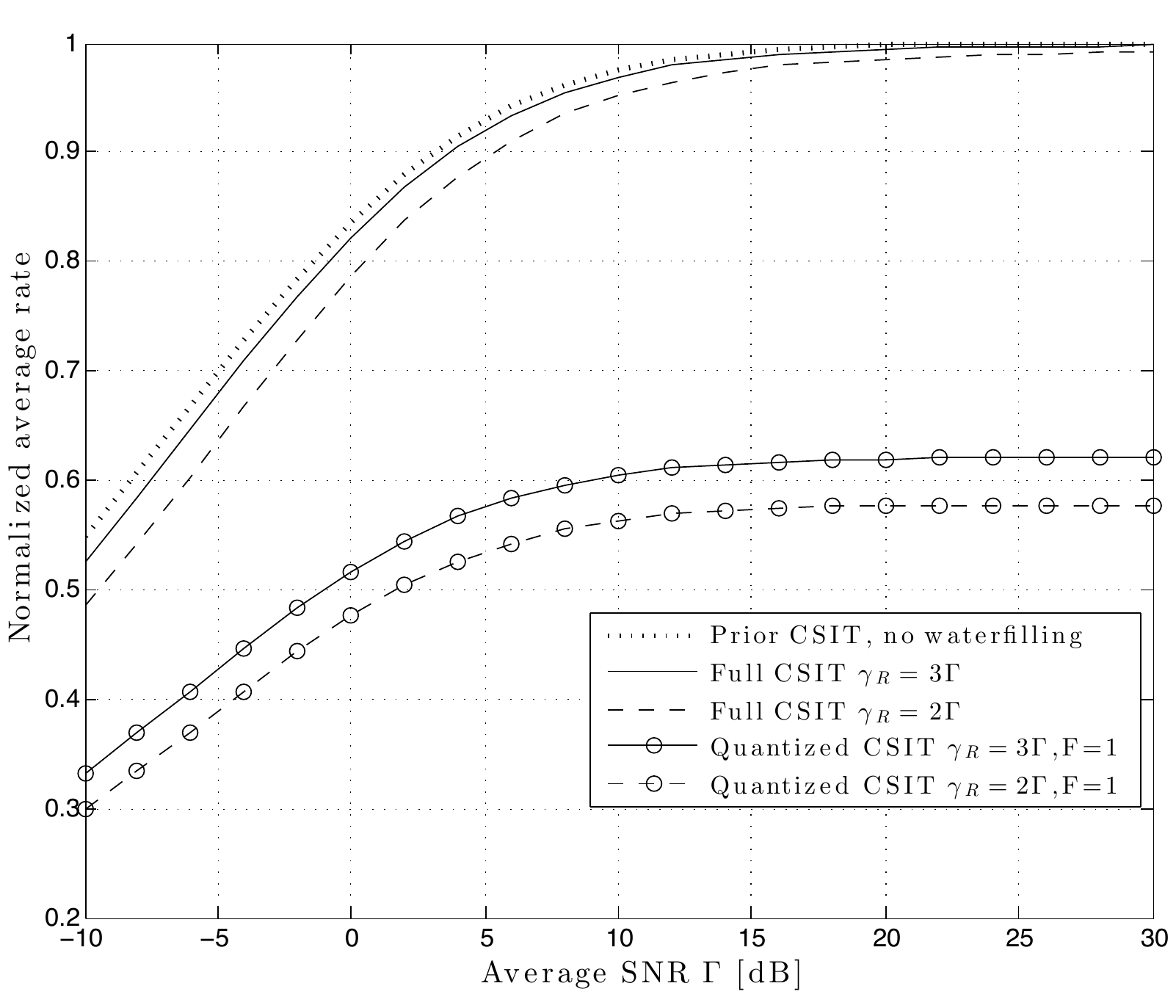}
\caption{Average rates as a function of the average SNR in a Rayleigh fading, normalized with the average rate achievable with prior CSIT and waterfilling. }\label{fig:Rate_vs_SINR}
\vspace{-12pt}
\end{figure}

We evaluate BRQ by assuming Rayleigh block fading, independent from slot to slot. Fig.~\ref{fig:Rate_vs_SINR} depicts the average rate (average throughput) of various schemes as a function of the mean SNR $\Gamma$. For each $\Gamma$, the average rate is normalized by the optimal average rate that can be achieved if CSIT is known a priori and water filling is applied. Two schemes with posterior CSIT are evaluated, full CSIT and quantized CSIT with $F=1$ bit per SNR, respectively. For each scheme, two different transmission rates are chosen, $R=\log_2(1+2\Gamma)$ and $R=\log_2(1+3\Gamma)$, respectively. Note that by such a choice, we are fixing the decoding probability to $p_R=e^{-2}$ and $p_R=e^{-3}$, respectively. The scheme with quantized CSIT is evaluated by using the rate distortion function of the exponential distribution, such that $d$ is determined according to~(\ref{eq:exp_dist_bound}), which is suboptimal. As a reference, we have also plotted the average rate when full CSIT is known a priori, but no water filling (fixed power) is applied. When full CSIT is available and the transmission rate $R$ is sufficiently high, then the knowledge of posterior CSIT is equally useful as the prior CSIT for average SNRs $\Gamma=10$ and higher. Furthermore, as water filling is more significant at low SNRs, we can see that with $R=\log_2(1+3\Gamma)$ and full CSIT, BRQ tightly approaches the average rate obtainable by water filling for $\Gamma$ equal to $20$ dB or higher. The scheme with quantized CSIT converges to a fixed value as $\Gamma$ increases (proof omitted due to lack of space).  

\begin{figure}[t]
  \centering
   \includegraphics[width=8.3cm]{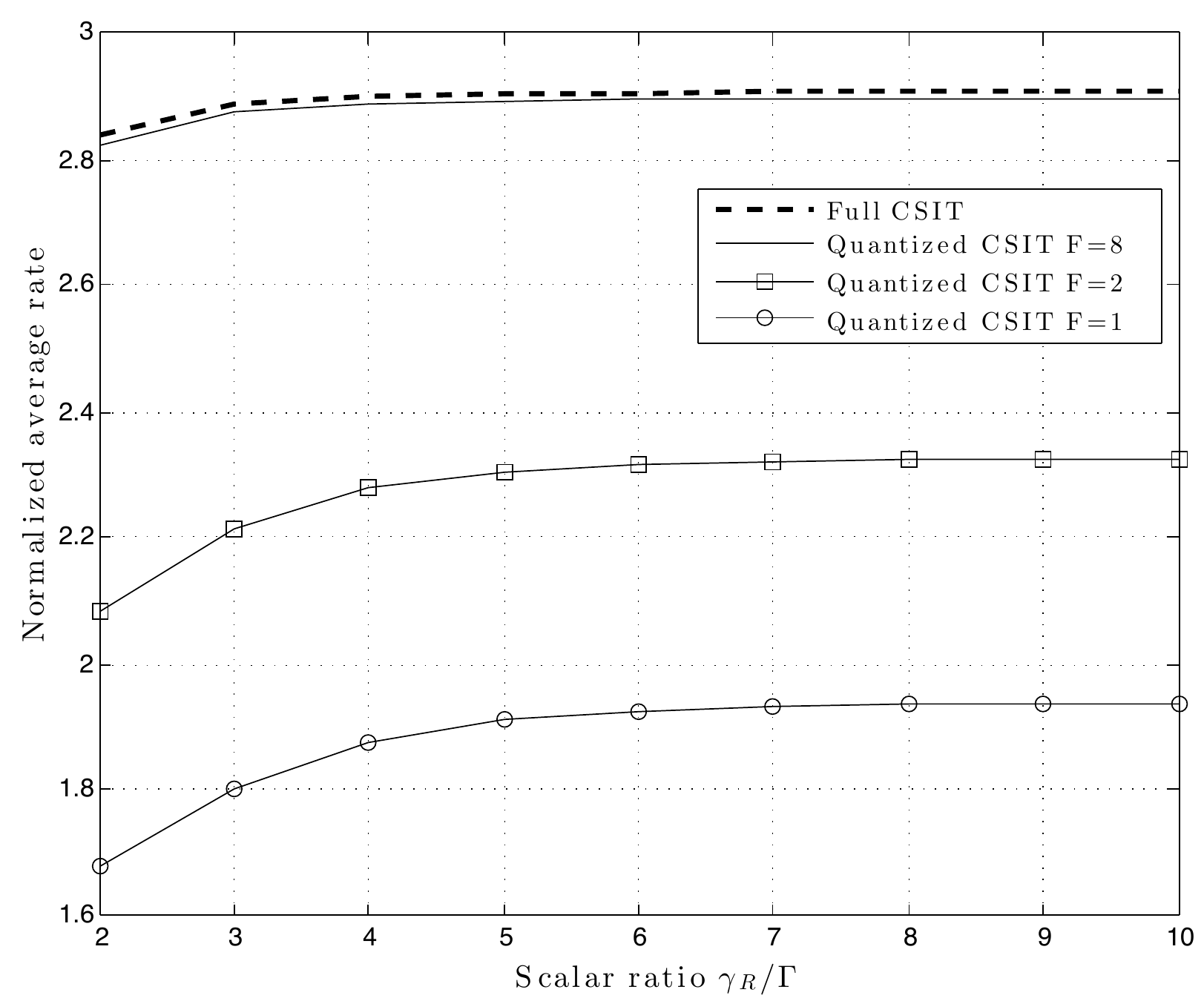}
\caption{Average rates as a function of the transmission rate $R$, represented through the equivalent SNR $\gamma_R$. The average SNR is fixed to $\Gamma=10$. }\label{fig:Rate_vs_gR}
\vspace{-12pt}
\end{figure}

The behavior of the average rate for the quantized scheme is better visible from~(\ref{fig:Rate_vs_gR}), where the average SNR is fixed to $\Gamma$, while $R=\log_2(1+\gamma_R)$ increases. The abscissa shows the scalar ratio $\frac{\gamma_R}{\Gamma}$, while the ordinate shows the absolute value of the average rate in bits per channel use [bit/c.u.]. We see that the schemes $F=1, F=2$ still grow in the region in which the scheme with $F=8$ is saturated. The reason is that the schemes with low $F$ are additionally affected by the decrease of $H(p_R)$ as $p_R$ grows, which affects the value of $d$, while $H(p_R)$ has a negligible effect for $F=8$.

\section{Discussion and Conclusions}
\label{sec:Discussion}

The main objective of this paper is to introduce a class of new transmission schemes based on backtrack retransmission (BRQ) that are useful when the CSIT can only be available after the transmission has taken place. Compared to the existing works on the topic, BRQ offers an elegant way to use the full posterior CSIT, based on adaptive Markov block coding, thereby closely approaching the average throughput achievable with prior CSIT. By extending the ideas of BRQ to the case where only a finite number of feedback bits are available, we have introduced a way to use the feedback bits which significantly departs from the standard way in which these bits are used in HARQ protocols. We devise a scheme in which it is possible to assemble multiple feedback bits and jointly report multiple CSIT values through vector quantization. 

In this initial work on the topic we have used the asymptotic information-theoretic results, valid when the number of channel uses per slot $N$ and when the number of slots in a block $L$ (used in vector quantization) goes to infinity. When finite values of $N$ and $L$ are considered, then the protocol should be adjusted in order to account for nonzero probability of error in the transmission as well as in the recovery based on the random binning. This analysis will be put in an extended version of this paper. Another interesting aspect is the choice of the distortion criterion. We have chosen $d$ to be constant, but in general $d$ should depends on the SNR value and the optimal $d$ should maximize the average rate. 

An interesting extension would be to generalize BRQ to multi-user scenarios. On the other hand, considering the practicality of the assumption about the posterior CSIT, the presented method is relevant for the practical wireless systems, such as LTE, based on fixed-size physical resources and significantly affected by intermittent interference from neighboring cells. It is therefore important to explore how BRQ can be implemented with practical coding/modulation schemes. 

\section*{Acknowledgement}
The author thanks Prof. Osvaldo Simeone from NJIT for his comments and pointing out the relation to Markov block coding; as well as to Kasper F. Trillingsgaard from Aalborg University for his comments and corrections. 

Part of this work has been performed in the framework of the FP7 project ICT-317669 METIS, which is partly funded by the European Union. The author would like to acknowledge the contributions of his colleagues in METIS, although the views expressed are those of the author and do not necessarily represent the project.


%
%
%
%
%
%
%
%
%

\end{document}